\documentclass[namedreferences]{kluwer}[1998/02/11]
\usepackage[dvips]{epsfig}
 
\begin{document}     
\begin{article}
\begin{opening}
\title{Searching for X-ray Luminous Starburst Galaxies}
\author{Andreas \surname{Zezas}\email{aze@star.le.ac.uk}}
\runningauthor{Zezas, Alonso-Herrero \& Ward}
\runningtitle{X-ray Luminous Starburst Galaxies}     
\institute{Department of Physics and Astronomy, University of Leicester,
Leicester, LE1 7RH, UK}
\author{Almudena \surname{Alonso-Herrero}\email{aalonso@as.arizona.edu}}  
\institute{Steward Observatory, the University of Arizona, Tucson, 
AZ 85721, USA}
\author{Martin J. \surname{Ward}\email{mjw@star.le.ac.uk}}
\institute{Department of Physics and Astronomy, University of Leicester,
Leicester, LE1 7RH, UK}

\begin{abstract}

The existence or otherwise of X-ray luminous star-forming galaxies
 has been an open
 question since the era of the Einstein satellite. Various 
 authors have claimed the discovery of X-ray luminous star-forming galaxies but
 in many cases more 
 careful spectroscopic studies of these objects have shown that many of
 them are in fact obscured AGN.
 In order to investigate the possibility that such
 a class of galaxies do exist, we have carried out a cross-correlation
 between optical and IRAS samples of galaxies which are known to contain large
 numbers
 of star-forming galaxies and catalogs of sources detected in X-ray
 surveys. The selection criteria for the optical
 follow-up observations was based on
 their X-ray and infrared (IRAS) colors and their X-ray
luminosities. We note that this sample is by no means complete or
 uniformly selected and hence cannot be used for statistical studies, but
 nevertheless confirmation of the existence of such a class of objects
 would be a important step, and would require us to understand the 
 physical process responsible for such powerful X-ray emission. 

 We have initiated an optical spectroscopic survey in order
 to obtain accurate spectroscopic classifications for all the objects
 which are claimed to be starburst galaxies. Here we present preliminary
 results from this survey. We have discovered a number of
 starburst galaxies with X-ray luminosities above
 $\sim10^{41}\,$erg s$^{-1}$ (for $\rm{H_{0}=50\,km\,s^{-1}\,Mpc^{-1}}$).
 We investigate possible
 origins for the X-ray emission in individual cases.

\end{abstract}
\keywords{\LaTeXe, (electronic) submission, user manual, class file, Kluwer}
\end{opening}   

\section{Introduction}

 It has been suggested by several authors (e.g., Bookbinder et al.
1980; Griffiths \& Padovani
1990; McHardy et al. 1998) that X-ray luminous 
($L_{\rm X}>10^{41-42}\,$erg s$^{-1}$ in the
$0.5-2.5\,$keV band for $H_0=50\,$km s$^{-1}$ Mpc$^{-1}$) starburst 
galaxies may make a significant contribution to the extragalactic
X-ray background (XRB). However there is still a considerable debate  
as to whether significant numbers of these objects  exist
(e.g., Moran et  al. 1994, 1996; Wisotzki \& Bade 1997).  Major
contributions are known to result from unobscured AGN, and reprocessed
emission from obscured AGN (Fabian et al. 1990).
Moran et al. (1994) defined the lower limit for the X-ray
luminosity of X-ray luminous star-forming galaxies at
$10^{42}\,$erg s$^{-1}$, however for our study, we will take 
$L_{\rm X}=10^{41}\,$erg s$^{-1}$ as the lower limit. From recent 
X-ray studies 
of nearby  galaxies (Read et al. 1997) the highest observed X-ray 
luminosity for starburst galaxies is
$\sim10^{41}\,$erg s$^{-1}$; the only exception is Arp~299 which 
has $L_{\rm X}=5 \times1 0^{41}\,$erg s$^{-1}$ in the $0.5-2.5\,$keV band 
(Zezas et al. 1998). 
A luminosity of $\sim10^{41}\,$erg s$^{-1}$ is still $10-100$ times
greater than the typical luminosity for nearby star-forming galaxies.

 In recent deep {\it ROSAT} X-ray surveys (e.g., Georgantopoulos et al. 1996;  
McHardy et al. 1998), a population of faint soft X-ray sources 
($f_{\rm X}\sim10^{-14}-10^{-15}\,$erg cm$^{-2}$ s$^{-1}$) with relatively 
high X-ray luminosities has been identified with narrow emission line
galaxies (NELGs). The
typical X-ray luminosities of these objects are about
$10^{42}\,$erg s$^{-1}$ or even higher. 
Modeling of the various classes of objects (QSOs, clusters, NELGs)  
which contribute to the
XRB (Pearson et al. 1997; McHardy et al. 1998) has shown that depending
on their evolution,  NELGs may 
contribute up to 50\% in the $0.5-2.0\,$keV range at fainter
fluxes  ($f_{\rm X}<10^{-15}\,$erg cm$^{-2}$ s$^{-1}$).
 Optical spectroscopy of these objects has shown that there is a significant
fraction of  bona-fide starburst galaxies together with a number of
objects lying on the borderline between starbursts and AGN (Griffiths et al. 
1996; McHardy et al. 1998). However, there has been 
a poor record in the past,  with many misclassified starbursts, which 
after a careful spectroscopic classification turned out to 
be AGN (e.g., Halpern et al. 1995; Moran et al. 1996).  Moreover, very 
recently Georgantopoulos et al. (1999) using the X-ray luminosity 
function of nearby starbursts and  AGN
estimated that the contribution of star-forming galaxies to the XRB is
less than $\sim10\%$. Indeed based on recent very deep {\it ROSAT} 
observations of
the Lockman Hole region (Hasinger et al. 1998) it has been claimed
that essentially all of the XRB can explained by various types of AGN.

 In order to obtain more insight into this problem and investigate 
the importance of
the class of X-ray luminous starbursts to the XRB, we have 
initiated an optical spectroscopic survey of a large sample of candidate 
for X-ray luminous starburst galaxies. We present preliminary results from
the first observing campaign and report the discovery of three 
galaxies belonging to this class.

\section{The Sample}

 Our sample was selected by cross-correlating suitable optically and
infrared selected lists of galaxies  with a potentially high content of
starburst galaxies, with the {\it ROSAT} X-ray catalogues.
 We have used the  Markarian (Mazzarella \& Balzano 1986) and Kiso-Schmidt 
(Takase \& Miyauchi-Isobe 1984) catalogues of UV-excess galaxies  and the
IRAS Faint Source Catalogue (IRAS FSC, Moshir 1991). 
These have  been cross-correlated
against the {\it ROSAT} bright source catalogue 
(RASSBSC, Voges et al. 1996) and the
WGAcat (White et al. 1995). The RASSBSC was constructed  after the
All Sky Survey carried-out in the first phase of the service of {\it ROSAT}.
It contains 18,811 X-ray sources down to a count rate of 0.5
$\rm{counts~s^{-1}}$ in the $0.1-2.4\,$keV band. 
However, because of the non-uniform coverage of the sky and the
effective sensitivity changes due to variations in the hydrogen
column density, this sample is neither uniform  nor complete.
The WGAcat  contains all the sources  detected in all the
{\it ROSAT} PSPC  pointed observations that became public prior to 1994 March.
The WGAcat is not uniform, but it contains much
fainter sources than the RASSBSC as the exposures of the pointed
observations are much deeper than the snapshots used to construct
the RASSBSC.
\begin{table}
\caption{The cross-correlations.}
\begin{tabular}{lccc}
\hline
Sample 1 & Sample 2 & Result & Chance-coincidences. \\
\hline
Markarian & RASSBSC & 26 & $0\pm0 $ \\
Markarian & WGAcat & 40 & $2\pm1 $  \\
\hline
IRAS FSC & RASSBSC & 1776 & $19\pm3 $ \\
IRAS FSC & WGAcat & 1514 & $66\pm5 $ \\
\hline
 Kiso & RASSBSC & 58 & $2\pm1$ \\
 Kiso & WGAcat & 76 & $5\pm1$ \\
\hline 
\end{tabular}
\end{table}

 We used a search radius of $30''$ which gives an association likelihood
greater than 95\% (the typical error radius in the 
WGAcat coordinates is $\sim20''$).  We assessed the chance coincidence 
probability by shifting the coordinates of one catalogue by a few
degrees and cross-correlating the catalogues again (this test was done
five times with different shifts to improve the statistics of the
chance coincidences).
The results from the cross-correlations together with the number of chance
coincidences are presented in Table~1.

From the cross-correlated lists we first excluded all the Galactic 
objects. We found  150, 126 and 447 X-ray emitting 
galaxies in the Markarian, Kiso lists and the IRAS FSC 
respectively. In order to
exclude AGNs  from the IRAS FSC sample we
applied a further selection criterion which uses the $L_{\rm X}/L_{\rm IR}$ 
ratio to distinguish 
between broad-line objects (Seyfert 1s and QSOs) and 
NLEGs (Seyfert 2s, LINERs and starbursts see Green et al. 
1992). After the application of this selection criterion, 
this catalogue was reduced to  a total of 310 entries. 
Finally, we excluded  all the known AGNs based on
classifications available in the literature. The result is a list of
$\sim100$  potential X-ray luminous star-forming galaxies.

%

\section{Observations and Activity Classification}

 We have initiated an optical spectroscopic survey of the final sample 
using the University of Arizona 2.3-m telescope on Kitt Peak. Observations  
were obtained during two observing runs in May and September 1998. 
Intermediate resolution  optical spectra covering the
wavelength range $3900-7100\,$\AA \ with a resolution of 
8\,\AA \   FWHM (which corresponds to $\simeq 400\,$km s$^{-1}$ at 5500\AA) 
have been obtained for 49 galaxies from our sample. We used a  
slit width of  $2.5''$  and the plate scale is  $1.6''$/pixel.

 The data have been reduced following the standard procedures 
(bias subtraction, flat-fielding and cosmic ray removal).
Standard stars from the list of 
Massey et al. (1988)  were used for both removal of the atmospheric 
absorption features and flux calibration. One-dimensional spectra with an 
optimum aperture were extracted from the two-dimensional data.

We have measured line fluxes and redshifts for all the galaxies observed 
in the first observing run. To assess the activity class, we made
use of the standard diagnostic line ratios 
[O\,{\sc iii}]$\lambda$5007/H$\beta$, 
[O\,{\sc i}]$\lambda$6300/H$\alpha$, 
[N\,{\sc ii}]$\lambda6584$/H$\alpha$,
and [S\,{\sc ii}]$\lambda\lambda$6713,6731/H$\alpha$ (e.g., Veilleux \&
Osterbrock 1987). Out of the 19 objects analyzed so far, we have found 
three X-ray luminous star-forming galaxies. These three galaxies lie well
within  the starburst locus of the diagnostic
diagrams.  The names of the X-ray luminous starburst 
galaxies together with their distances\footnote{we report the 
first measurement for the redshift of  
IRAS~F16419+8213} are presented in Table~2. The optical line 
ratios and the visual extinctions derived from the Balmer decrement
are also given in Table~2.

The X-ray fluxes are calculated from the point spread function
(PSF)  and vignetting corrected
WGAcat count rates. We used a Raymond \& Smith (1977) thermal plasma
model with a temperature $kT=0.8\,$keV, which is found to adequately
describe 
the soft X-ray  ($0.5-2.5\,$keV)  spectra of star-forming  galaxies
(Read et al.  1997). For the absorbing column density we used the Galactic
from 21\,cm radio observations
(Stark et al. 1992).

\begin{figure}
\centerline{\epsfig{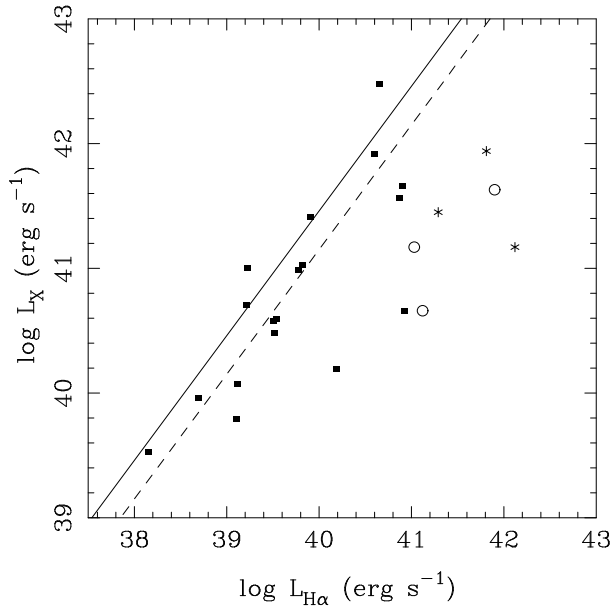}}
\vspace{0.5cm}
\caption{Low luminosity end of the soft ($0.5-2.5\,$keV) X-ray luminosity 
versus the
$\rm{H\,{\sc \alpha}}$ luminosity correlation (the two lines represent
different fits to the correlation for X-ray 
luminosities ranging from $10^{40}\rm{erg\,s^{-1}}$ to 
$10^{46}\rm{erg\,s^{-1}}$, Koratkar et al. 1995) found for  AGN 
(squares). The star symbols are our new X-ray luminous starbursts. For
comparison we also plot three nearby luminous starforming galaxies
(open circles).}
\end{figure}

\section{Results}
In Table~3 we give the H$\alpha$ luminosity (corrected for extinction), 
the FIR luminosity ($42.5-112.5\,\mu$m band) estimated using the 
expression of David et al. (1992) and the 
soft X-ray luminosity for the three X-ray luminous starbursts. We can obtain 
an estimate  of the star formation rate (SFR) for 
massive stars ($M > 10\,$M$_\odot$) using the calibration
in terms of the H$\alpha$ luminosity given in Kennicutt (1983).  
Somewhat surprisingly the derived SFRs are moderate, comparable to
the values found for nearby starburst galaxies.


\begin{table}
\footnotesize
\caption{Spectroscopic properties of the galaxies.}
\begin{tabular}{lcccccc}
\hline
Galaxy & vel & [N\,{\sc ii}]/H$\alpha$ &  
[S\,{\sc ii}]/H$\alpha$ 
& [O\,{\sc i}]/H$\alpha$ & [O\,{\sc iii}]/H$\beta$   & $A_V$ \\
 (1) & (2) & (3) & (4) & (5) & (6) & (7) \\
\hline
NGC~6013  & 4508 &  0.49 & 0.38 & 0.04 & 1.78 & 3.0  \\
IRAS~F16400+3944 & 8934 & 0.35 & 0.35 & 0.05 & 0.63 & 2.6  \\
IRAS~F16419+8213 &  11993 & 0.28 & 0.27 & 0.12 & 0.68 & 2.6 \\
\hline
\end{tabular}
Columns~(2) is the recession velocity in $\rm{km\,s^{-1}}$. 
Columns~(3), (4), (5) and (6) are the observed optical line
ratios. Column~(7) is the visual extinction.
\end{table}

Figure~1 shows  the low luminosity end of the soft X-ray versus the
$\rm{H\,{\sc \alpha}}$  correlation found for AGN (filled
squares, Koratkar et al. 1995). We plot our X-ray luminous starbursts 
as star symbols and other nearby starbursts as open
circles (Read et al. 1997; Zezas et al. 1998; 
Alonso-Herrero et al. 1999). It is clear that 
the three X-ray luminous galaxies present an 
excess of H$\alpha$ emission with respect to the X-ray emission 
when compared with low-luminosity AGNs, but they seem to be 
similar to the nearby X-ray luminous starbursts.  If we estimate 
the X-ray luminosities from the
number of massive X-ray binaries (using the 
total number of ionizing photons obtained from the 
extinction-corrected $\rm{H\,{\sc \alpha}}$ luminosity, or
the FIR luminosity) we find that the observed X-ray luminosities
cannot be reproduced.  
The estimated X-ray luminosities are low by a factor of $\sim10-100$  
even if we add 
a contribution from a superwind (e.g., Heckman et al. 1995).
 This discrepancy may be due either to the presence of a highly obscured AGN
component or another mechanism which makes the production of
X-rays from a starburst more efficient (e.g., subsolar metalicities
Griffiths \& Padovani 1990).

\begin{table}[h]
\caption{Luminosities and SFRs for the X-ray luminous starbursts.}
\begin{tabular}{lccccc}
\hline
Galaxy &  $L_{{\rm H}\alpha}$ & $L_{\rm FIR}$ &
$L_{\rm X}$  & Total SFR   \\
       & ($10^{41}\,{\rm erg\,s}^{-1}$) & ($\rm{10^{43}\,erg\,s^{-1}}$) & 
($\rm{10^{41}erg\,s^{-1}}$) & ($\rm{ M_{\odot}\,yr^{-1}})$  \\
 (1)  & (2) & (3) & (4) & (5) &  \\
\hline
NGC~6013  & 2.0 & 2.5 & 2.8 & 0.2 \\
IRAS~F16400+3944  & 13.2  & 30.4 & 1.5 & 2.0 \\
IRAS~F16419+8213  & 6.6 & 9.5 & 8.9 & 1.0 \\
\hline
\end{tabular}

Column~(2) is the extinction-corrected H$\alpha$ luminosity.  
Column~(3) is the FIR luminosity. Column~(4) is
the star fomation rate (for stars with $M > 10\,$M$_\odot$)
from $\rm{H\,{\sc \alpha}}$ luminosity. Column~(5) is 
the $0.5-2.5\,$keV X-ray luminosity.
\end{table}

\section{Conclusions}

 We present preliminary results from an on-going spectroscopic
survey which has the main goal of identifying examples of
 X-ray luminous star-forming galaxies. Our results suggest that
 such galaxies DO exist in the local universe. This may have
important implications for the study of the extragalactic X-ray
background and for understanding the mechanisms responsible for their
high X-ray luminosity.
Future observations will enlarge the current sample and help answer 
the question first raised by Bookbinder et al. (1980) more than
15 ago years, of whether X-ray
luminous starburst galaxies make a significant contribution to the XRB.

\begin{acknowledgements}

AA-H acknowledges support from the National
Aeronautics and Space Administration on grant NAG 5-3042 through the
University of Arizona. 

\end{acknowledgements}

\end{article}
\end{document}